\documentclass[english,prd, 11pt, nofootinbib,  superscriptaddress, preprintnumbers,floatfix]{revtex4-2}
\usepackage[T1]{fontenc}
\usepackage{babel}
\usepackage{mathrsfs}
\usepackage{bm,amsmath}
\usepackage{amssymb}

\usepackage{graphicx}
\usepackage{hyperref}
\hypersetup{pdftex,colorlinks=true,linkcolor=blue,citecolor=blue,menucolor=black,urlcolor=blue,filecolor=blue}

\usepackage{slashed}
\usepackage{bbold}
\usepackage{multirow}

\usepackage{amsfonts}
\usepackage{makecell}
\usepackage[dvipsnames]{xcolor}
\usepackage[normalem]{ulem}
\usepackage{longtable}
\usepackage{array}
\usepackage{float}
\usepackage{subfigure}

\newcommand{\dsz}{D_{s0}^*(2317)}
\newcommand{\dso}{D_{s1}(2460)}
\newcommand{\dspipi}{D_s^+\pi^+\pi^-}

\newcommand*{\Rom}[1]{\uppercase\expandafter{\romannumeral #1\relax}}

\newcommand{\itp}{\affiliation{CAS Key Laboratory of Theoretical Physics, Institute of Theoretical Physics,\\
Chinese Academy of Sciences, Beijing 100190, China}}

\newcommand{\ucas}{\affiliation{School of Physical Sciences, University of Chinese Academy of Sciences, Beijing 100049, China}}

\newcommand{\peng}{\affiliation{Peng Huanwu Collaborative Center for Research and Education, Beihang University, Beijing 100191, China}}

\newcommand{\hiskp}{\affiliation{Helmholtz-Institut f\"ur Strahlen- und Kernphysik and Bethe Center for Theoretical Physics,\\ Universit\"at Bonn, D-53115 Bonn, Germany}}

\newcommand{\fzj}{\affiliation{Institute for Advanced Simulation and Institut f\"ur Kernphysik, Forschungszentrum J\"ulich, D-52425 J\"ulich, Germany}}

\newcommand{\tbilisi}{\affiliation{Tbilisi State University, 0186 Tbilisi, Georgia}}

\begin{document}
\title{Isospin-conserving hadronic decay of the ${D_{s1}(2460)}$ into ${D_s\pi^+\pi^-}$}
\author{Meng-Na Tang}\email{tangmengna@itp.ac.cn}\itp\ucas

\author{Yong-Hui Lin}\email{yonghui@hiskp.uni-bonn.de}\hiskp

\author{Feng-Kun Guo}\email{fkguo@itp.ac.cn}\itp\ucas\peng

\author{Christoph Hanhart}\email{c.hanhart@fz-juelich.de}\fzj

\author{Ulf-G.~Mei{\ss}ner}\email{meissner@hiskp.uni-bonn.de}\hiskp\fzj\tbilisi

\begin{abstract}
The internal structure of the charm-strange mesons $D_{s0}^*(2317)$ and $D_{s1}(2460)$ are subject of intensive studies.
Their widths are small because they decay dominantly through isospin-breaking hadronic channels $D_{s0}^*(2317)^+\to D_s^+\pi^0$ and $D_{s1}(2460)^+\to D_s^{*+}\pi^0$.
The $D_{s1}(2460)$ can also decay into the hadronic final states $D_s^+\pi\pi$, conserving isospin. In that case there is, however, a strong suppression from
phase space.
We study the transition $D_{s1}(2460)^+\to D_s^+\pi^+\pi^-$ in the scenario that the $D_{s1}(2460)$ is a $D^*K$ hadronic molecule. The $\pi\pi$ final state interaction is taken into account through dispersion relations. 
We find that the ratio of the partial widths of the $\Gamma(D_{s1}(2460)^+\to D_s^+\pi^+\pi^-)/\Gamma(D_{s1}(2460)^+\to D_s^{*+}\pi^0)$ obtained in the
molecular picture is consistent with the existing experimental measurement.
More interestingly, we demonstrate that the $\pi^+\pi^-$ invariant mass distribution shows a double bump structure,
which can be used to disentangle the hadronic molecular picture from the compact state picture for the $D_{s1}(2460)^+$.
Predictions on the $B_{s1}^0\to B_s^0\pi^+\pi^-$ are also made.
\end{abstract}
\maketitle

\section{Introduction}

The study of exotic hadrons with heavy quarks commenced with the discovery of the scalar charm-strange meson $\dsz$ decaying to $D_s^+\pi^0$ by the BaBar Collaboration~\cite{BaBar:2003oey} and the axial-vector charm-strange meson $\dso$ decaying to $D_s^{*+}\pi^0$ by the CLEO Collaboration~\cite{CLEO:2003ggt}. In fact, in the {BaBar} data of the $D_s^+\pi^0\gamma$ invariant mass distribution with $D_s^+\gamma$ constrained in the $D_s^{*+}$ signal region, there is also a peak around 2.46~GeV~\cite{BaBar:2003oey}, which could correspond to the $\dso$ state. No isospin partners for these states have been found, and their widths are extremely small, with upper bounds of 3.8~MeV and 3.5~MeV for the $\dsz$ and $\dso$, respectively~\cite{ParticleDataGroup:2022pth}. Thus, these two mesons are isoscalar states. Since their masses are much lower than the quark model predictions of the lowest $c\bar s$ mesons with the corresponding $J^{P}$ quantum numbers~\cite{Godfrey:1985xj}, various models were proposed to understand them, including, for instance, modifying the $c\bar s$ quark model~\cite{Cahn:2003cw}, interpreting the $\dsz$ and $\dso$ as $D^{(*)}K$
hadronic molecules, respectively~\cite{Barnes:2003dj,vanBeveren:2003kd,Kolomeitsev:2003ac,Chen:2004dy,Guo:2006fu,Guo:2006rp}, 
compact tetraquarks~\cite{Maiani:2004vq,Wang:2006uba}, 
and chiral partners of the ground state $D_s$ and $D_s^*$ mesons~\cite{Bardeen:2003kt,Nowak:2003ra}.
{Tremendous progress has been made towards understanding the $D_{s0}^*$ and $D_{s1}$, as well as their nonstrange partners, using lattice quantum chromodynamics or by analyzing the lattice data~\cite{Liu:2012zya,Mohler:2012na,Mohler:2013rwa,Altenbuchinger:2013vwa,Lang:2014yfa,MartinezTorres:2014kpc,Guo:2015dha,Moir:2016srx,Bali:2017pdv,Guo:2018kno,Cheung:2020mql,Gayer:2021xzv,Gregory:2021rgy,Lang:2022elg,Yang:2021tvc} (for a recent review, see Ref.~\cite{Guo:2023wkv}).
Important information on the internal structure of these mesons can also be obtained from $B_{(s)}$ decays~\cite{Albaladejo:2016hae,Du:2017zvv,Du:2019oki,Liu:2022dmm} and $e^+e^-$ collisions~\cite{Wu:2022xzh}. }

Crucial observables to distinguish the hadronic molecular scenario from the others are the isospin breaking hadronic decay widths
$\dsz \to D_s^+\pi^0$ and $\dso \to D_s^{*+}\pi^0$, which are of the order of 100~keV for hadronic
molecules~\cite{Faessler:2007gv,Liu:2012zya,Guo:2018kno} and much smaller in the other models~\cite{Godfrey:2003kg,Colangelo:2003vg,Bardeen:2003kt}. 
The reason is that as $D^{(*)}K$ hadronic molecule, the $\dsz$ ($\dso$) strongly couples to $D^{(*)}K$ and the isospin splittings of the charged and neutral $D^{(*)}$ and $K$ mesons lead to significant isospin breaking effects, since the respective poles are
located rather close to the thresholds.
Radiative decays of the $\dsz$ and $\dso$ have been computed in Refs.~\cite{Faessler:2007gv,Gamermann:2007bm,Lutz:2007sk,Cleven:2014oka,Fu:2021wde} in the hadronic molecular model and in Ref.~\cite{Bardeen:2003kt} in the chiral doublet model.

For the $\dso$, in addition to the isospin-breaking hadronic decay into the $D_s^+\pi^0$, also the decay into $D_s^{+} \pi^{+} \pi^{-}$
is allowed kinematically, respecting isospin symmetry, since the two pions can be in even partial waves and thus in an isoscalar state. 
The ratio of the partial width of this three-body decay relative to the two-body hadronic decay has been measured by the Belle Collaboration as~\cite{Belle:2003kup}
\begin{equation}
	\frac{\Gamma\left(D_{s 1}(2460)^{+} \rightarrow D_s^{+} \pi^{+} \pi^{-} \right)}{\Gamma\left(D_{s 1}(2460)^{+} \rightarrow D_s^{*+} \pi^0\right)}= 0.14 \pm 0.04 \pm 0.02 , \label{eq:ratioexp}
\end{equation}
while the value from the fit by the Particle Data Group (PDG) is $0.09\pm0.02$~\cite{ParticleDataGroup:2022pth}.
Not much work has been done regarding the decay $D_{s 1}(2460)^{+} \rightarrow D_s^{+} \pi^{+} \pi^{-}$. 
In  Ref.~\cite{Fajfer:2015zma}, by treating the $\dso$ as a $P$-wave charm-strange meson, the width of the $D_{s 1}(2460)^{+} \rightarrow D_s^{+} \pi^{+} \pi^{-}$ was predicted to be about  0.25~keV. In that work the outgoing $D_s$ is taken at rest such that the 
pion pair must be in a $P$-wave to conserve parity and angular momentum. 
Accordingly, in that work also the two--pion decay is isospin violating in contrast
 to our calculation, where the $P$-wave sits between the outgoing $D_s$ and the isoscalar pion pair.

The latest calculation of the width of $D_{s 1}(2460)^{+} \rightarrow D_s^{*+} \pi^{0}$ within the $D^*K$ molecule scenario for $D_{s 1}(2460)$ is given in Ref.~\cite{Fu:2021wde} where its value of $(111\pm 15)~{\rm keV}$ is obtained from the complete isospin breaking contributions in the framework of unitarized chiral perturbation theory (UChPT) up to the next-to-leading order.
In this work we explicitly calculate the two-pion transitions and demonstrate that, assisted with the result
of Ref.~\cite{Fu:2021wde},
the  ratio in Eq.~\eqref{eq:ratioexp} is consistent with the $D^*K$ molecular picture for the $D_{s 1}(2460)$. 
We
also show the  internal structure of  the $D_{s 1}(2460)$
leaves a characteristic imprint on
 the $\pi^+\pi^-$ invariant mass distribution in the decay $D_{s 1}(2460)^{+} \rightarrow D_s^{+} \pi^{+} \pi^{-}$.

Furthermore, predictions on the $B_{s1}\to B_s^0\pi^+\pi^-$ will be made, where the $B_{s1}$ is the bottom partner of the $\dso$.

\section{Decay of the \texorpdfstring{$\bm{D_{s 1}(2460)^{+}}$}{Ds1(2460)} as a hadronic molecule into \texorpdfstring{$\bm{D_s^{+} \pi^{+} \pi^{-}}$}{Ds pi+ pi- }}

In this section, we calculate the decay width and $\pi^+\pi^-$ invariant mass distribution of the $\dso\to\dspipi$, taking into account the $S$-wave final-state interaction (FSI) between the two pions.

A crucial quantity distinguishing a hadronic molecular state from a compact state is the coupling of the state to the constituent hadrons,
because the coupling squared is proportional to the probability of the physical state being composite~\cite{Weinberg:1965zz,Baru:2003qq,Guo:2017jvc}. Here we focus on the $\dso$. If the $\dso$ is a purely compact state with negligible $D^*K$ component, then its coupling to $D^*K$ would be negligibly small. In contrast, if the $\dso$ is a pure $D^*K$ bound state, then its coupling to $D^*K$ is maximal and the corresponding
loops appear in all transitions at leading order, some times accompanied by short-ranged operators to absorb the
pertinent divergences.
The three-body decay into $\dspipi$ can proceed through the diagrams shown as Fig.~\ref{fig:diagram}~(a), (b) and (c).
The loops are divergent and accordingly there is necessarily a counter term, shown as diagram (d), at the same order. 
In the molecular picture the effect of this counter term can be estimated as $(\gamma/\beta)=25\%$, where
$\beta=0.77$~GeV is the mass of the lightest exchange particle allowed (in this case the $\rho$ meson) and
$\gamma=\sqrt{2\mu \epsilon}=0.19$~GeV is the binding momentum, with $\mu$ for the reduced mass of the $D^*K$ system and
$\epsilon=45$~MeV for the binding energy.\footnote{Note that this estimate is build on the
concept of resonance saturation which requires to employ a natural cut-off in the calculation as
we do it below~\cite{Epelbaum:2001fm}.} If on the other hand the $\dso$ were a compact state,
the loops would be strongly suppressed and the transition amplitude would be dominated by diagram~(d).

\subsection{\texorpdfstring{$\bm{D_{s 1}(2460)^{+} \to D_s^{+} \pi^{+} \pi^{-}}$}{Ds1 to Ds pi pi} through the \texorpdfstring{$\bm{D^*K}$}{D* K} component}
\begin{figure}[t]
	\centering
	\includegraphics*[width=0.8\textwidth,angle=0]{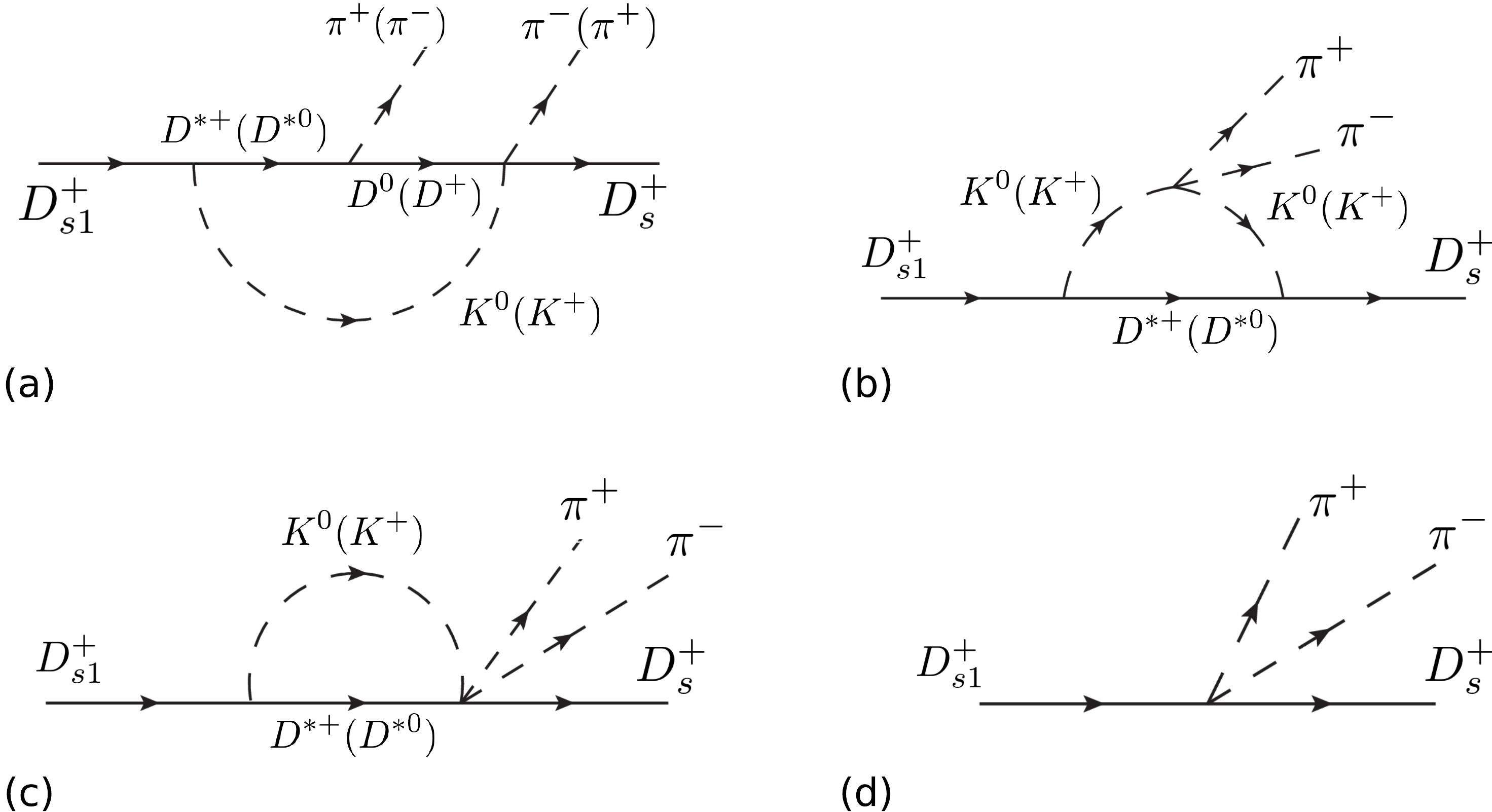}
	\caption{
		Diagrams for the decay $D_{s 1}(2460)^{+} \to D_s^{+} \pi^{+} \pi^{-}$ with (a+b+c) and without (d) the $D^*K$ contribution.
	}
	\label{fig:diagram}
\end{figure} 
Let us consider first the decay of the $D_{s 1}(2460)^{+} \to D_s^{+} \pi^{+} \pi^{-}$ with the molecular assumption occurring through the  one-loop triangle diagrams (a), (b) and (c) in Fig.~\ref{fig:diagram}. 
For the decay $\dso\to\dspipi$ to keep isospin symmetry, the $\pi^+\pi^-$ system in the final-state must be an isospin scalar. Therefore, the quantum numbers of the $\pi^+\pi^-$ system must be $J^{PC}={\rm even}^{++}$. Then the lowest partial wave between the $\pi^+\pi^-$ system and $D_s^{+}$ is a $P$-wave.

To calculate the amplitudes for diagrams (a), (b) and (c) in Fig.~\ref{fig:diagram}, we employ 
the following effective Lagrangians for the $D_{s 1} D^* K$~\cite{Cleven:2014oka} and the other vertices~\cite{Wise:1992hn,Yan:1992gz},
\begin{align}\label{eq:lagrangian}
	{{\cal L}_{P D^{*\dagger} D_{s1}}}&=\frac{f}{\sqrt{2}} D_{s 1}^\mu\left( D^{*+\dagger}_{\mu} K^{0\dagger}+D^{*0\dagger}_{\mu} K^{+\dagger}\right)+{\rm h.c.}, \\
	{{\cal L}_{\chi}}&=\frac{F_\pi^2}{4}{\rm Tr}\left[\partial_\mu U \partial^\mu U ^\dagger\right]- \left\langle H_a({\rm i} v\cdot {\cal D}_{ab})\bar{H}_b \right\rangle +g \left\langle H_a\gamma\cdot{\cal A}_{ab}\gamma_5\bar{H}_{b}\right\rangle, \label{eq:lagrangian2}
\end{align}
where $U =u^2=\exp(\sqrt{2}{\rm i}\Phi/F_\pi)$ is a nonlinear function of the field $\Phi$ for the light pseudoscalar Goldstone bosons with 
\begin{equation}
	\Phi = \left(\begin{array}{ccc}
	\frac{\pi^0}{\sqrt{2}}+\frac{\eta}{\sqrt{6}} & \pi^{+} & K^{+} \\
	\pi^{-} & -\frac{\pi^0}{\sqrt{2}}+\frac{\eta}{\sqrt{6}} & K^0 \\
	K^{-} & \bar{K}^0 & -\frac{2 \eta}{\sqrt{6}}
	\end{array}\right),
\end{equation}
with $F_\pi$ the pion decay constant in the chiral limit.
$H_a=1/2(1+v\cdot\gamma)[P_{a,\mu}^*\gamma^\mu-P_a\gamma_5]$, with $a$ the light flavor index, is a superfield for the ground state pseudoscalar and vector heavy mesons, which are in the same heavy quark spin multiplet, where 
$P^*_\mu$ and $P$ annihilate the vector and pseudoscalar heavy mesons, respectively. 
${\cal D}_{ab}^\mu=\delta_{ab}\partial^\mu-{\cal V}_{ab}^\mu$ is the chirally covariant derivative with ${\cal V}_\mu=(u^\dagger\partial_\mu u+u \partial_\mu u^\dagger)/2$ the light meson vector current and ${\cal A}_\mu={\rm i}(u^\dagger\partial_\mu u-u \partial_\mu u^\dagger)/2$ the corresponding axial current. Tr$[\cdot]$ and $\left\langle\cdot\right\rangle$ take traces in the flavor and spinor spaces, respectively. 

Since the $\dso$ mass is smaller than the $D^*K$ threshold by just about 45~MeV, the binding momentum of the $\dso$ as a $D^*K$ bound state is about 0.19~GeV, much smaller than both the kaon and $D^*$ masses. 
Thus, 
we may use a constant coupling for the $D_{s 1} D^* K$ coupling in Eq.~\eqref{eq:lagrangian}, following Ref.~\cite{Cleven:2014oka}. The coupling $f$ squared can be computed from the residue of the unitarized $D^*K\to D^*K$ scattering amplitude in UChPT, and it is related to the $\dsz DK$ coupling by means of heavy quark spin symmetry as done in Ref.~\cite{Fu:2021wde}. We take  $f=10.1^{+0.8}_{-0.9}$~GeV from Ref.~\cite{Fu:2021wde}, which is the result in UChPT using the low-energy constants (LECs) determined in Ref.~\cite{Liu:2012zya}.
The axial coupling constant $g$ is determined to be $0.565\pm0.006$ by reproducing the measured partial width of $D^{*+}\to D^0\pi^+$, that is, $(83.4\pm 1.8)\ {\rm keV}$ for the total width of $D^{*+}$ with the branching fraction $0.677\pm0.005$~\cite{ParticleDataGroup:2022pth}. For the pion decay constant, we use the physical value $F_\pi=92~{\rm MeV}$.

The amplitude for the triangle diagrams in Fig.~\ref{fig:diagram} is given by
\begin{equation}\label{eq:sce1}
{{\cal M}_\text{(a)+(b)+(c)}}={\cal M}_\text{(a)}+{\cal M}_\text{(b)}+{\cal M}_\text{(c)}+{\cal M}_\text{(a)}|_{p_{\pi^+}\leftrightarrow p_{\pi^-}}+{\cal M}_\text{(b)}|_{p_{\pi^+}\leftrightarrow p_{\pi^-}}+{\cal M}_\text{(c)}|_{p_{\pi^+}\leftrightarrow p_{\pi^-}},
\end{equation}
where ${\cal M}_\text{(a)}$, ${\cal M}_\text{(b)}$ and ${\cal M}_\text{(c)}$ read
\begin{align}\label{eq:Ma}
{{\cal M}_\text{(a)}}=&\,\frac{-f g M_D\sqrt{M_{D_s}M_{D^*}}}{2 F_\pi^3}\int\frac{{\rm d} ^4 k}{(2\pi)^4}
\frac{{\rm i} v\cdot(k-p_{D_{s 1}}-p_{\pi^-})\, \epsilon_{D_{s1}}^{(\lambda)}\cdot p_{\pi^+}}{(k^2-M_{D^*}^2)[(p_{D_{s 1}}-k)^2-M_{K}^2][(k-p_{\pi^+})^2-M_D^2]},\\
{{\cal M}_\text{(b)}}=&\,\frac{f g M_D\sqrt{M_{D_s}M_D^*}}{12 F_\pi^3}\int\frac{{\rm d}^4 k}{(2\pi)^4}
\frac{-2 {\rm i} \epsilon_{D_{s1}}^{(\lambda)}\cdot (p_{D_{s 1}}-p_{\pi^+}-p_{\pi^-}-k)}{(k^2-M_{D^*}^2)[(p_{D_{s 1}}-k)^2-M_{K}^2][(p_{D_{s 1}}-p_{\pi^+}-p_{\pi^-}-k)^2-M_K^2]}\notag\\
&\times \bigg(M_{D_{s 1}}^2-M_\pi^2+p_{D_{s 1}}\cdot (2p_{\pi^-}-4p_{\pi^+}-2k)+k^2-2p_{\pi^+}\cdot p_{\pi^-}+k\cdot(4p_{\pi^+}-2p_{\pi^-})\bigg),\label{eq:Mc}\\
 \label{eq:Mb}
{{\cal M}_\text{(c)}}=&\,\frac{-{\rm i} f g \sqrt{M_{D_s}M_{D^*}}}{12 F_\pi^3}\int\frac{{\rm d}^4 k}{(2\pi)^4}
\frac{\epsilon_{D_{s1}}^{(\lambda)}\cdot (p_{\pi^+}-k-2p_{\pi^-})}{(k^2-M_K^2)[(p_{D_{s 1}}^{}-k)^2-M_{D^*}^2]}~.
\end{align}
Here, 
$\epsilon_{D_{s1}}^{(\lambda)}$ is the polarization vector of the $D_{s1}$, with $\lambda$ denoting the polarization components, $p_{D_{s1}}$ is the four-momentum of the $D_{s1}$ meson, and $p_{\pi^\pm}$ is the four-momentum of the $\pi^\pm$ emitted from the $D_{s1}$ decay.

In the UChPT calculation for the isospin-breaking hadronic decays and radiative decays of the $\dso$ and $\dsz$, a three-momentum cutoff $q_{\rm max}=745^{+35}_{-37}~{\rm MeV}$ is introduced~\cite{Fu:2021wde}.
Here, we use the same cutoff range for the loop integrals to ensure the treatment to be consistent with the calculation of the decay  $D_{s 1}(2460)^{+} \rightarrow D_s^{*+} \pi^{0}$ in Ref.~\cite{Fu:2021wde}.

\subsection{Partial wave projection and the \texorpdfstring{$\bm{\pi\pi}$}{pi pi} FSI}
\begin{figure}[t]
	\centering
	\includegraphics*[width=0.5\textwidth,angle=0]{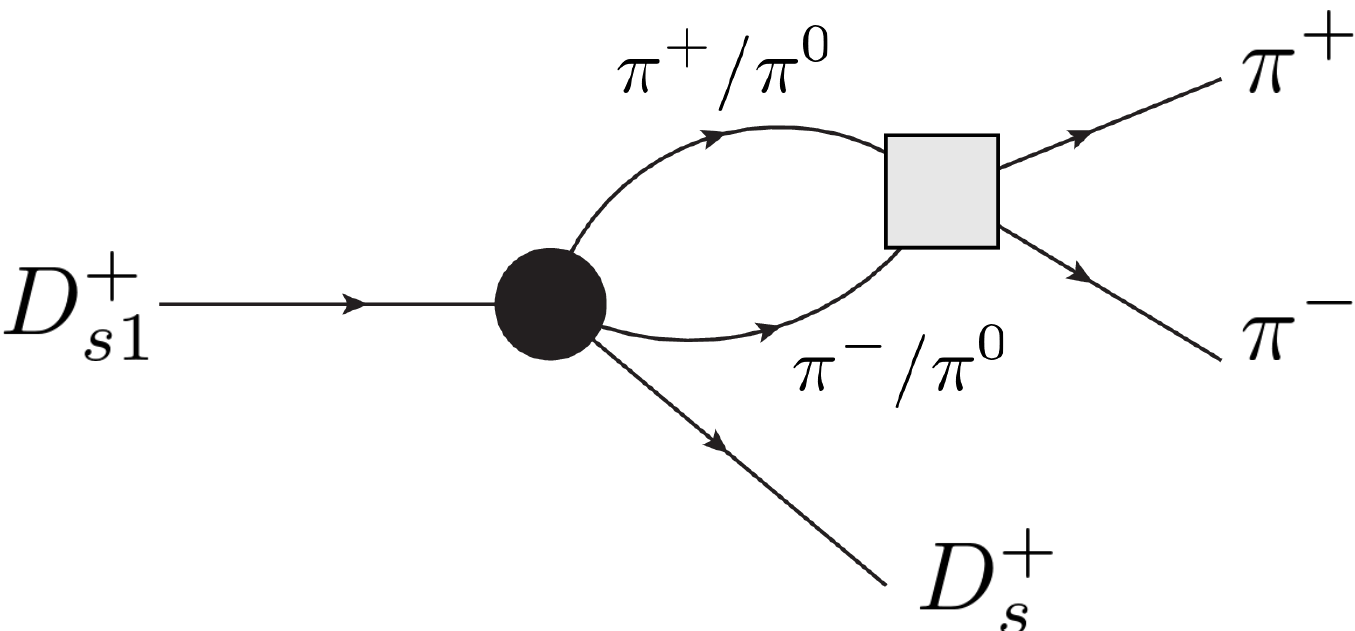}
	\caption{
		Diagram for the decay $D_{s 1}(2460)^{+} \to D_s^{+} \pi^{+} \pi^{-}$ with the $\pi\pi$ FSI considered. The black circle denotes the amplitude of $D_{s 1}(2460)^{+} \to D_s^{+} \pi^{+} \pi^{-}$ for the diagrams in Fig.~\ref{fig:diagram} and similar loop diagram contributions to $D_{s 1}(2460)^{+} \to D_s^{+} \pi^{0} \pi^{0}$. The square represents the pion-pion rescattering.  
	}
	\label{fig:fsi}
\end{figure} 

Since the two pions can be in the isoscalar $S$-wave and the phase space allows the $\pi^+\pi^-$ invariant mass to be up to 0.49~GeV, the $\pi\pi$ final state interaction (FSI) needs to be considered  in the calculation of both the $\dso\to\dspipi$ partial width and the corresponding $\pi^+\pi^-$ invariant mass distribution.
In particular, the $f_0(500)$ resonance, also known as the $\sigma$ meson, contributes through the $S$-wave $\pi\pi$ FSI.

To that end, we briefly introduce the partial wave projection that is used to include the $S$-wave $\pi\pi$ FSI effect. We work in the rest frame of $D_{s 1}(2460)^{+}$ and choose the positive $z$-axis to be along the moving direction of $D_s^+$. The decay amplitude ${\cal M}$ of $D_{s 1}(2460)^{+} \to D_s^{+} \pi^{+} \pi^{-}$ in Eq.~\eqref{eq:sce1} is a function of the polarization vector $\epsilon_{D_{s 1}}^{(\lambda)}$ of $D_{s 1}$ and two kinematic variables $m_{\pi^+\pi^-}$ and $\theta$, i.e., ${\cal M}={\cal M}(m_{\pi^+\pi^-}, \theta, \epsilon_{D_{s 1}}^{(\lambda)})$, where $m_{\pi^+\pi^-}$ is the $\pi^+\pi^-$ invariant mass and $\theta$ denotes the angle between the $\pi^+$ moving direction in the center-of-mass (c.m.) frame of $\pi^+\pi^-$ and the moving direction of the $\pi^+\pi^-$ system in the $D_{s1}$ rest frame.  Then the partial wave projection is carried out by means of the formula (see, e.g., Ref.~\cite{Gulmez:2016scm})
\begin{align}\label{eq:pwa}
	{\cal M}^{l}(m_{\pi^+\pi^-})&= 
 \frac{2\pi {\rm Y}_{\bar{l}}^0(\hat{\mathrm{\mathbf{z}}})}{2J+1}\sum_{\substack{\lambda, m}}\int {\rm d}\cos\theta\, {\rm Y}_l^m(\theta)^*
 (m 0 m|l S J)(0 \lambda m|\bar{l}\bar{S}J) {\cal M}(m_{\pi^+\pi^-}, \theta, \epsilon_{D_{s 1}}^{(\lambda)}),
\end{align}
where $(mS_zJ_z|lSJ)$ are the Clebsch-Gordan coefficients for the coupling of orbital angular momentum $l$ and spin $S$ to the total angular momentum $J$, with $m,S_z$ and $J_z$ the corresponding third components, and ${\rm Y}_l^m(\theta)$ is the spherical harmonic function. 
Here, we use $l,S$ and $\bar l,\bar S$ to denote the quantum numbers of the $\pi^+\pi^-$ and $D_{s1}(2460)D_{s}$ systems, respectively.
Thus, for the $S$-wave $\pi\pi$, we have $\{J=0,l=S=0,\bar{l}=\bar{S}=1\}$; for the $D$-wave $\pi\pi$, we have $\{J=l=2,S=0,\bar{l}=\bar{S}=1\}$ and $\{J=l=2,S=0,\bar{l}=3,\bar{S}=1\}$. Since the lightest tensor meson $f_2(1270)$ is far away from the region of interest, only the $S$-wave $\pi\pi$ FSI effect is taken into account.

The $\pi\pi$ FSI for the decay $D_{s 1}(2460)^{+} \to D_s^{+} \pi^{+} \pi^{-}$, as shown in Fig.~\ref{fig:fsi}, is taken into account using a dispersion relation approach with inhomogeneity including the left-hand cut contribution. A similar approach has been used in studying other hadronic processes, see., e.g., Refs.~\cite{Anisovich:1996tx,Garcia-Martin:2010kyn,Kubis:2015sga,Kang:2013jaa,Chen:2015jgl,Chen:2016mjn,Danilkin:2020kce,Baru:2020ywb}. 
Here, the inhomogeneity comes from the loop diagrams shown in Fig.~\ref{fig:diagram} (a), (b) and (c).
The decay amplitude ${\cal M}_{\pi^+\pi^-}$ for the process $D_{s 1}(2460)^{+} \to D_s^{+} \pi^{+} \pi^{-}$ with the $S$-wave $\pi\pi$ rescattering {(here only the $\pi^0\pi^0$ and $\pi^+\pi^-$ channels are relevant due to the phase space limitation, and the $D$-wave rescattering is negligible because the $f_2(1270)$ mass is much higher than the energy region of interest)} included satisfies the unitarity relation 
\begin{equation}
	\label{eq:unitarity}
        \frac{1}{2{\rm i}}{\rm disc}\, {\cal M}_{\pi^+\pi^-}^{00}=\frac23 (t^0_0)^*\sigma_\pi {\cal M}_{\pi^+\pi^-}^{00}+\frac13 (t^0_0)^*\sigma_\pi {\cal M}_{\pi^0\pi^0}^{00}=(t^0_0)^*\sigma_\pi {\cal M}_{\pi^+\pi^-}^{00},
\end{equation}
where ${\rm disc}\, {\cal M}_{\pi^+\pi^-}^{00}$ is the discontinuity of the $D_{s 1}(2460)^{+} \to D_s^{+} \pi^{+} \pi^{-}$ amplitude in the isoscalar $S$ partial wave along the {$\pi\pi$} right-hand cut, starting from the $\pi\pi$ threshold to infinity along the positive real axis, and ${\cal M}_{\pi^0\pi^0}^{00}$ is the decay amplitude of $D_{s 1}(2460)^{+} \to D_s^{+} \pi^{0} \pi^{0}$ in the isoscalar $S$ partial wave. For the second identity, ${\cal M}_{\pi^0\pi^0}^{00}={\cal M}_{\pi^+\pi^-}^{00}$ in the isospin limit is applied. Furthermore, $\sigma_\pi=\sqrt{1-4M_\pi^2/s}$ is the phase space factor of the 
two-pion channel with $\sqrt{s}=m_{\pi\pi}$ the c.m. energy of the $\pi\pi$ system, $t_0^0$ represents the elastic isoscalar $S$-wave $\pi\pi$ scattering amplitude, and 
\begin{equation}
    (t_0^0)^*\sigma_\pi = e^{-{\rm i}\delta_0^0} \sin\delta_0^0,
\end{equation}
with $\delta_0^0$ the isoscalar $S$-wave $\pi\pi$ scattering phase shift. 
The Omn\`{e}s function $\Omega(s)$ is introduced as the solution of the unitarity relation
\begin{equation}
 	\label{eq:uni-omnes}
 	\frac{1}{2{\rm i}}{\rm disc}\, \Omega=(t^0_0)^*\sigma_\pi \Omega,
\end{equation}
which can be solved by~\cite{Omnes:1958hv}
\begin{equation}\label{eq:omnes}
	\Omega(s)=\exp\left(\frac{s}{\pi}\int_{4M_\pi^2}^{\infty}\frac{{\rm d} s^\prime}{s^\prime}\frac{\delta_0^0(s^\prime)}{s^\prime-s-{\rm i} \epsilon}\right).
\end{equation}
Starting from Eq.~\eqref{eq:unitarity} and Eq.~\eqref{eq:uni-omnes}, one can obtain
\begin{equation}
	\label{eq:uni-fsi}
	\frac{1}{2{\rm i}}{\rm disc}\, \frac{{\cal M}^{00}-{\cal M}_{ L}^{00}}{\Omega}=\frac1\Omega(t^0_0)^*\sigma_\pi {\cal M}_{ L}^{00}=\frac1{|\Omega|}\sin \delta^0_0 {\cal M}_{ L}^{00},
\end{equation}
where ${\cal M}_L^{00}$ is the part of ${\cal M}_{\pi^+\pi^-}^{00}$ containing only the inhomogeneity that is modeled by diagrams (a)+(b)+(c) in Fig.~\ref{fig:diagram} in the present work, i.e., the $S$-wave projection of ${\cal M}_\text{(a)+(b)+(c)}$ in Eq.~\eqref{eq:sce1}. 

Then one can write a once-subtracted dispersion relation for the decay amplitude of $D_{s 1}(2460)^{+} \to D_s^{+} \pi^{+} \pi^{-}$ with the $S$-wave $\pi\pi$ FSI effect included,
\begin{equation}\label{eq:fsi-K}
	{\cal M}^{00}(s) = {{\cal M}_{L}^{00}} + \Omega(s)  \left[ a +
	\frac{s}{\pi}\int\limits_{4M_\pi^2}^{s_{\rm max}} \, \frac{{\rm d}s'}{s'} \, 
	\frac{{\cal M}_{L}^{00}\sin\delta_0^0(s')}{{\vert\Omega(s')\vert} \, (s'-s-{\rm i} \epsilon)} \right],
\end{equation}
where $a$ is a subtraction constant. 
A single subtraction is sufficient to ensure the convergence as can be checked by varying the cutoff
$s_{\rm max}$. 
Each term on the right-hand side of Eq.~\eqref{eq:fsi-K} can be interpreted diagrammatically. ${\cal M}_L^{00}$ is the amplitude of diagrams (a), (b) and (c) in Fig.~\ref{fig:diagram} projected to the $\pi\pi$ $S$ wave, as mentioned above. The second term in the square bracket (together with the $\Omega(s)$ factor outside) corresponds to the triangle diagrams connected to the $\pi\pi$ FSI. The subtraction term $a\,\Omega(s)$ corresponds to a $D_{s1} D_s \pi\pi$ contact term connected to the $\pi\pi$ FSI, which can be expressed as
\begin{equation}\label{eq:fsi-P}
	{\cal M}^{ 00}_{\rm comp.}(s) = g_c\,\epsilon_{D_{s1}} \cdot p_{D_s} \Omega(s) ,
\end{equation}
where $g_c$ is a $D_{s1}\to D_s\pi^+\pi^-$ contact term coupling constant.

As for the $\pi\pi$ scattering phase shift, we employ the following parameterization~\cite{Garcia-Martin:2011iqs},
\begin{equation}
	\label{eq:delta}
	\delta_0^0(s)=
	\begin{cases}
		0, &  0 \le \sqrt{s}\le 2M_\pi,\\
		{f_1}(s) , & 2M_\pi < \sqrt{s}\le \sqrt{s_m},
	\end{cases}
\end{equation}
where 
\begin{equation}
    {f_1}(s)=\operatorname{arccot}\left\{ \frac{s}{\lambda(s,M_\pi^2,M_\pi^2)^{1/2}}\frac{M_\pi^2}{s-z_0^2/2}\left[\frac{z_0^2}{M_\pi\sqrt{s}}+B_0+B_1 w(s)+B_2 w^2(s)+B_3 w^3(s)\right]\right\} . \label{eq:deltapara}
\end{equation}
The parameters of this expression,
$${z_0}=M_\pi,\quad B_0=7.14, \quad B_1=-25.3, \quad B_2=-33.2,\quad B_3=-26.2  ,$$
were adjusted to the $\pi\pi$ scattering phase shifts extracted 
from a Roy-type analysis of the two-pion system,
with 
\begin{equation}
	w(s)=\frac{\sqrt{s}-\sqrt{4 M_K^2-s}}{\sqrt{s}+\sqrt{4 M_K^2-s}}
\end{equation}
and $\lambda(x,y,z)=x^2+y^2+z^2-2(xy+yz+zx)$  is the K\"{a}ll\'{e}n function. 
Since here the phase space restricts the physical region of $\sqrt{s}$ to be less than 0.5~GeV, we only consider
contributions from the $\pi\pi$ channel to the rescattering. Accordingly, 
the phase shift $\delta_0^0$ is smoothly extrapolated from the matching point $\sqrt{s_m}=0.85~{\rm GeV}$ to the asymptotic value of $180^\circ$  at $\sqrt{s} = \infty$ following the prescription in Ref.~\cite{Moussallam:1999aq},
\begin{equation}
\delta_0^0(s) = \pi+[{f_1}(s_m)-\pi]\frac2{1+\left({s}/{s_m}\right)^{3/2}}, \qquad \sqrt{s_m} < \sqrt{s}.
\end{equation}

\section{Results}\label{sec:results}

\subsection{Results for \texorpdfstring{$\bm{\dso\to D_s^+\pi^+\pi^-}$}{Ds1 -> Ds+ pi+ pi-}}
\label{subsec:c}

To compare our results to  the data we consider two schemes that refer to different treatments of the
coupling $g_c$ of the contact term, see Eq.~\eqref{eq:fsi-P}.
Since in the hadronic molecular picture of the $\dso$, the compact contribution to its decay widths is expected to be relatively small, 
we set by hand  $g_c$ to zero in scheme~I. 
In scheme~II, the value of $g_c$ will be adjusted to reproduce the measured ratio in Eq.~\eqref{eq:ratioexp}
to check if its size is consistent with the naturalness estimate of 25\% provided above.

\begin{itemize}
	\item In scheme~I, the partial width of the $D_{s 1}(2460)^{+} \to D_s^{+} \pi^{+} \pi^{-}$ is determined to be 
	\begin{equation}
		\Gamma(D_{s 1}(2460)^{+} \to D_s^{+} \pi^{+} \pi^{-}) = (21\pm4)~{\rm keV}.
	\end{equation}
	In the numerical calculation, $s_{\rm max}$ has been set to $3~{\rm GeV}$, and we checked that varying $s_{\rm max}$ to even larger values for the dispersive integral in Eq.~\eqref{eq:fsi-K} leads to a change of the width by less than 5\%, much smaller than the uncertainty quoted above.
	The uncertainty comes from two sources: 
	varying the coupling constant $f$ within the range obtained in UChPT in Ref.~\cite{Fu:2021wde}; 
	varying the momentum cutoff of the triangle loop integrals $q_{\rm max}$ within $745^{+35}_{-37}~{\rm MeV}$, which is the range determined in Ref.~\cite{Fu:2021wde}. 
	The former and the latter sources contribute about 70\% and 30\%, respectively, to the uncertainty.
	
	Taking $\Gamma(D_{s 1}(2460)^{+} \to D_s^{*+} \pi^{0}) = (111\pm15)~{\rm keV}$ in the hadronic molecular model computed in the UChPT framework with the same LECs~\cite{Fu:2021wde}, we obtain 
	\begin{equation}
		\frac{\Gamma\left(D_{s 1}(2460)^{+} \rightarrow D_s^{+} \pi^{+} \pi^{-} \right)}{\Gamma\left(D_{s 1}(2460)^{+}  \rightarrow D_s^{*+} \pi^0\right)}\Bigg|_\text{mol.}= 0.19^{+0.07}_{-0.05} , \label{eq:ratioth}
	\end{equation}
	which is consistent with the Belle measurement given in Eq.~\eqref{eq:ratioexp} and with the PDG fit value within two sigma~\cite{ParticleDataGroup:2022pth}. 

	\item In scheme~II,  the subtraction constant in Eq.~\eqref{eq:fsi-K}, or equivalently the $g_c$ coupling in Eq.~\eqref{eq:fsi-P}, is adjusted to reproduce the experimental ratio of Eq.~\eqref{eq:ratioexp}. 
	Using the same range of $q_{\rm max} = 745^{+35}_{-37}~{\rm MeV}$ and $s_{\rm max}=3~{\rm GeV}$ as used in scheme~I, We obtain
	\begin{equation}
		g_c=2.1_{-2.0}^{+1.2}{}_{-1.4}^{+1.5}~{\rm GeV^{-1}},
		\label{eq:gc}
	\end{equation}
	where the first error comes from the uncertainties from the inputs, which include the experimental ratio in Eq.~\eqref{eq:ratioexp} and the width of $\Gamma(D_{s 1}(2460)^{+} \to D_s^{*+} \pi^{0})$ from Ref.~\cite{Fu:2021wde}, while the second error comes from the calculation performed in this work by varying the coupling constant $f$ and $q_{\rm max}$ within the ranges given in Ref.~\cite{Fu:2021wde}. 
	One finds that even a  value of $g_c $ equal to zero is consistent with the currently available data, 
 which means that the short-distance contribution, when the momentum cutoff $q_{\rm max}$ in restricted within the range given in Ref.~\cite{Fu:2021wde},
 is marginal. 
	With the value of $g_c$ given above, 
	the partial decay width of $D_{s 1}(2460)^{+} \to D_s^{+} \pi^{+} \pi^{-}$ is 
	\begin{equation}
		\Gamma(D_{s 1}^+ \to D_s^{+} \pi^{+} \pi^{-}) = \left(16_{-5}^{+7}\right)~{\rm keV}.
		\label{eq:widthc}
	\end{equation}
The difference of the central values of this equation and that of Eq.~(\ref{eq:ratioth}) is of the expected size and we therefore conclude that
the presently available data are fully consistent with a molecular nature of the $\dso$. 
\end{itemize}

\begin{figure}[t]
	\centering
	\includegraphics*[width=0.48\textwidth,angle=0]{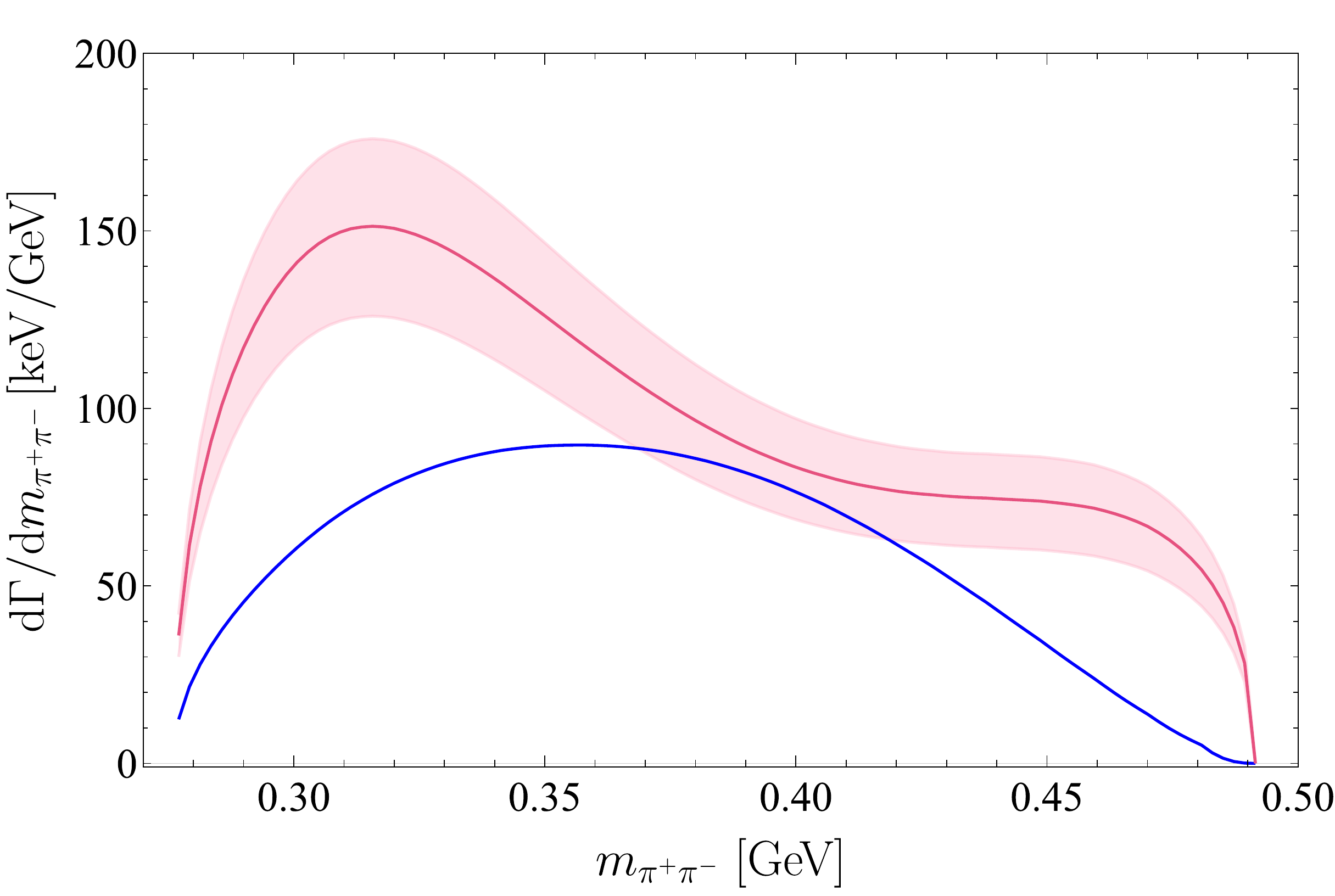}~~~~
	\includegraphics*[width=0.48\textwidth,angle=0]{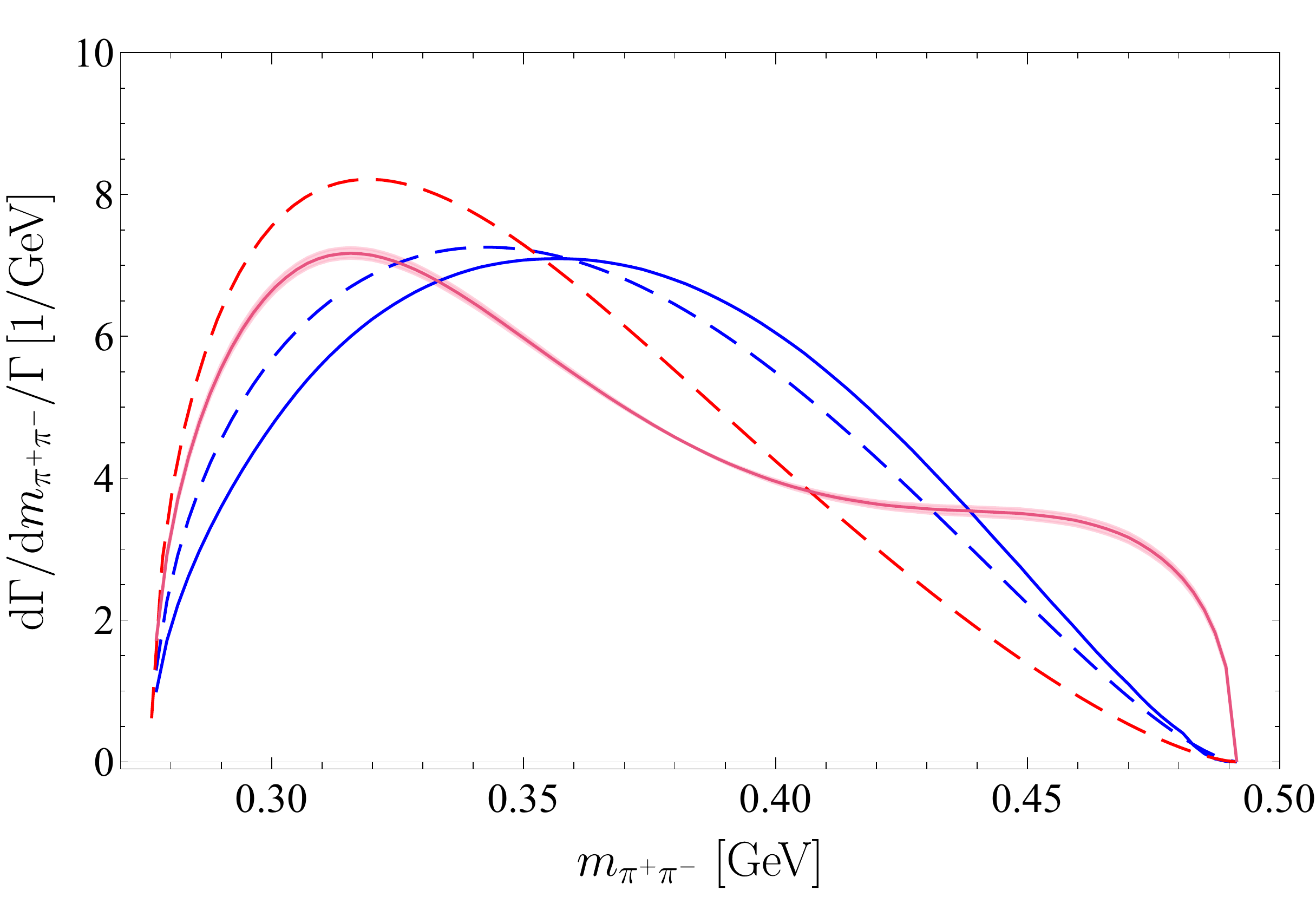}
	\caption{
		Results in scheme~I. Left panel: invariant mass distributions of $\pi^+\pi^-$ for the decay of $D_{s 1}(2460)^{+} \to D_s^{+} \pi^{+} \pi^{-}$. Right panel: the invariant mass distributions normalized  to the corresponding widths. 
		The red solid curves denote the results considering the loop diagrams in Fig.~\ref{fig:diagram} (a), (b) and (c) with the $S$-wave $\pi\pi$ FSI. The red dashed line in the right panel corresponds to the one without the FSI effect. 
		The light-red bands are the corresponding theoretical uncertainties propagated from those of the parameters in scheme~I. 
		For comparison, the blue solid and dashed lines are the results in the compact state model for the $\dso$, i.e., Fig.~\ref{fig:diagram}~(d) with and without the FSI included, respectively. {Dashed lines are only present in the right panel.}
	}
	\label{fig:IMD-err}
\end{figure}

\begin{figure}[tb]
	\centering
	\includegraphics*[width=0.48\textwidth,angle=0]{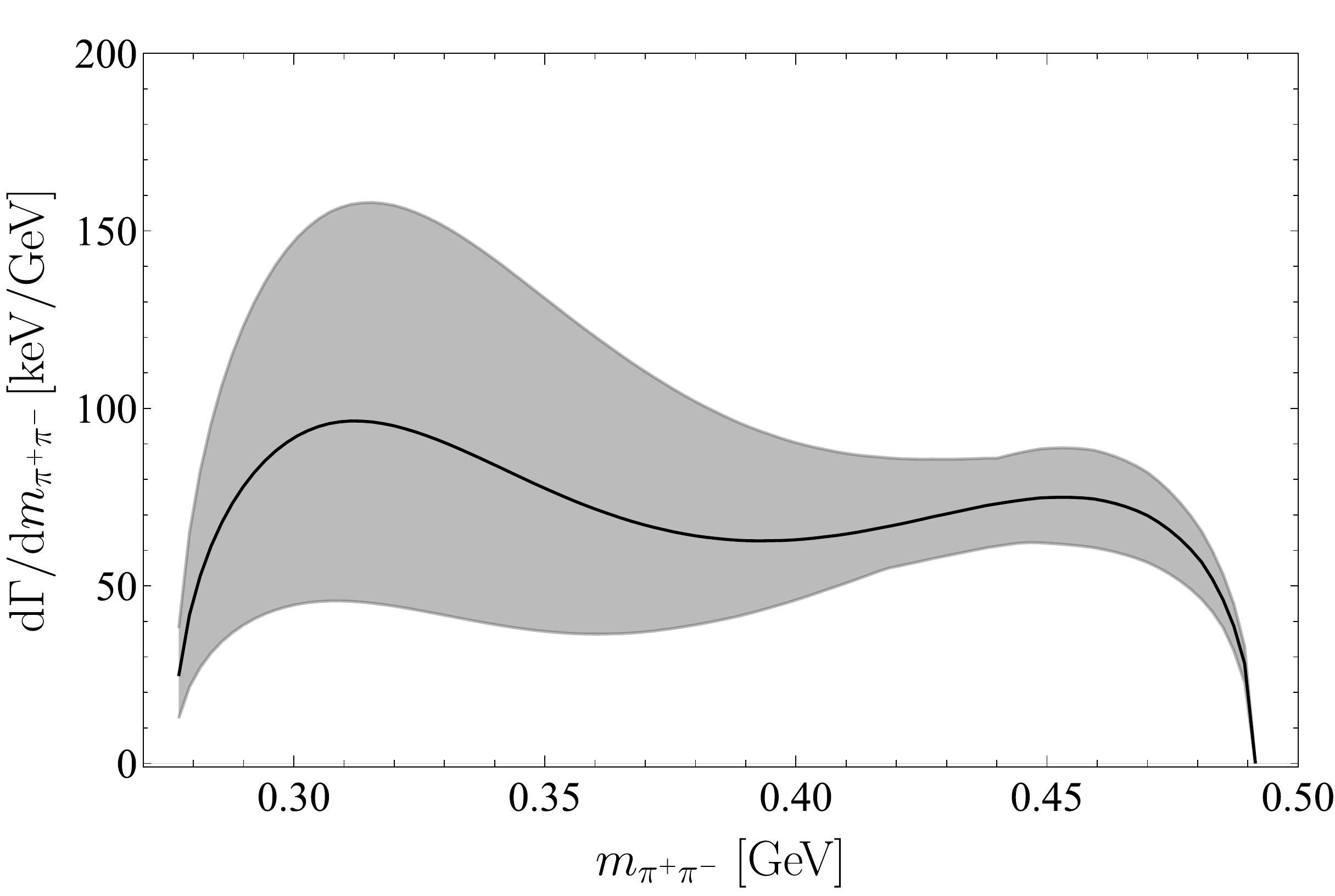}~~~~
	\includegraphics*[width=0.48\textwidth,angle=0]{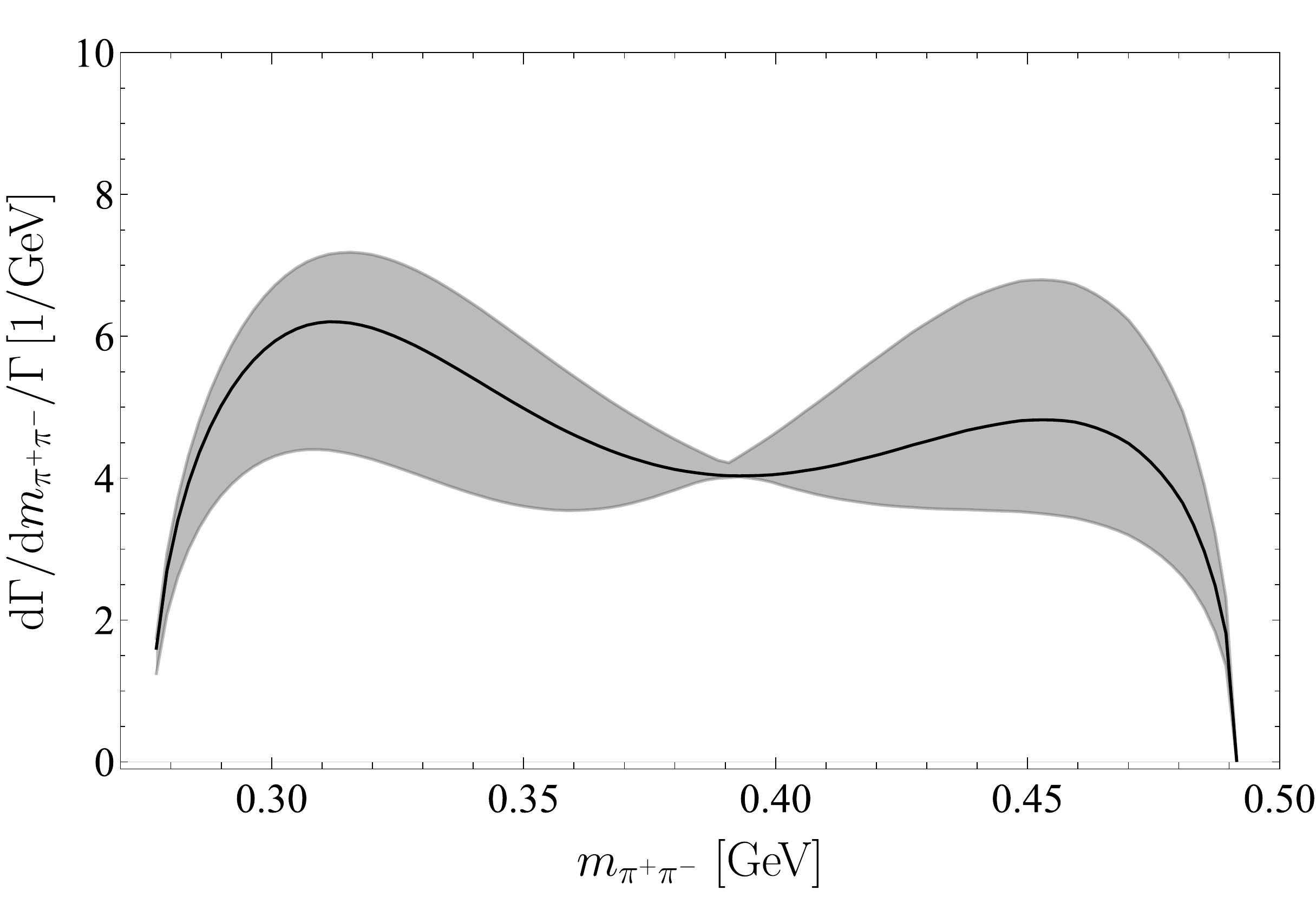}
	\caption{
		Results in scheme~II. Left panel: invariant mass distributions of $\pi^+\pi^-$ for the decay of $D_{s 1}(2460)^{+} \to D_s^{+} \pi^{+} \pi^{-}$. Right panel: the invariant mass distributions normalized  to the corresponding widths. 
		The black solid curves denote the results by adjusting the subtraction to reproduce the measured ratio in Eq.~\eqref{eq:ratioexp}.
		The bands are the corresponding theoretical uncertainties propagated from those of the parameters in scheme~II as detailed in the text. 
	}
	\label{fig:IMD-err2}
\end{figure}

Moreover, we find that the $\pi^+\pi^-$ invariant mass distribution can be used as an observable to distinguish the hadronic molecular approach from the compact state model for the $\dso$. 
In Figs.~\ref{fig:IMD-err} and \ref{fig:IMD-err2}, we show the $\pi^+\pi^-$ invariant mass distributions in scheme~I (the red curves and bands) and scheme~II, respectively.
One finds a double bump structure.
Such structure has two sources: the loop diagrams in Fig.~\ref{fig:diagram} and the $\pi\pi$ FSI.
To see this, we show as the red dashed curve in the right panel of Fig.~\ref{fig:IMD-err} the differential decay width of $D_{s 1}(2460)^{+} \to D_s^{+} \pi^{+} \pi^{-}$, divided by the integrated partial width, from the one-loop diagrams in Fig.~\ref{fig:diagram} without the $\pi\pi$ FSI. It peaks at around 0.31~MeV, just above the $\pi\pi$ threshold.
The $\pi\pi$ FSI then enhances the higher end of the $\pi\pi$ invariant mass distribution due to the existence of the $f_0(500)$ resonance whose information is contained in the $\pi\pi$ scattering phase shift $\delta_0^0$.

In contrast, if the $\dso$ is a compact state which does not couple to $D^*K$, the $\pi\pi$ invariant mass distribution would not receive contributions from the loop diagrams~(a), (b) and (c) in Fig.~\ref{fig:diagram}.
Then with $g_c=7~{\rm GeV^{-1}}$ that is adjusted to produce $13~{\rm keV}$ for the partial width of $D_{s 1}(2460)^{+} \to D_s^{+} \pi^{+} \pi^{-}$ using Eq.~\eqref{eq:fsi-P}, one obtains the blue solid curves in Fig.~\ref{fig:IMD-err}.
The distribution has a broader single bump and takes its maximum at around 0.36~GeV.
Switching off the $\pi\pi$ FSI in this case leads to the normalized differential distribution shown as the blue dashed curve in the right panel of Fig.~\ref{fig:IMD-err}.
One sees that the $\pi\pi$ FSI shifts the maximum of the bump to a higher energy.

Therefore, a high statistics measurement of the $\pi\pi$ invariant mass distribution from
$D_{s 1}(2460)^{+} \to D_s^{+} \pi^{+} \pi^{-}$  would provide one with direct access
to the molecular component of the $\dso$.

\subsection{Predictions on \texorpdfstring{$\bm{B_{s1}^0\to B_s^0\pi^+\pi^-}$}{Bs1 -> Bs pi+ pi-}}
\label{subsec:b}

Employing heavy quark flavor symmetry, we can make predictions on the decay $B_{s1}\to B_s^0\pi^+\pi^-$, where $B_{s1}$ is the bottom partner of the $\dso$.
The $B_{s1}$ has been predicted in the hadronic molecular model as an isoscalar $B^*\bar K$ bound state~\cite{Kolomeitsev:2003ac,Guo:2006rp,Albaladejo:2016lbb,Du:2017zvv,Fu:2021wde,Yang:2022vdb}.
The predicted mass $(5774 \pm 13)$~MeV~\cite{Fu:2021wde} is in  accordance with the lattice result on the lowest $B_{s1}$ meson $(5750 \pm 17 \pm 19)~\mathrm{MeV}$~\cite{Lang:2015hza} and slightly larger than the more recent lattice determination of $(5741 \pm 14)~\mathrm{MeV}$~\cite{Hudspith:2023loy}.
So far, the only experimentally observed $B_{s1}$ meson is the $B_{s 1}(5830)^0$~\cite{ParticleDataGroup:2022pth}, which is the bottom partner of the $D_{s1}(2536)$ and not of the one discussed here.

Here, we predict the partial width of the $B_{s 1}^{0} \to B_s^{0} \pi^{+} \pi^{-}$ and the corresponding $\pi^+\pi^-$ invariant mass distribution. The $B_{s1}$ is treated as a $B^*\bar K$ molecular state, and the framework is the same as that for the $D_{s1}$ in Section~\ref{subsec:c}.
We take scheme~II with the contact term fixed from reproducing the measured ratio in the charm sector in Eq.~\eqref{eq:ratioexp}.
Heavy quark flavor symmetry requires the contact term coupling in the bottom sector to take a value given by that in Eq.~\eqref{eq:gc} multiplied by $\sqrt{M_{B_{s1}}M_{B_s}/(M_{D_{s1}}M_{D_s})}$. 
The $B_{s1}B^*\bar K$ coupling is related to that of the $D_{s1}D^*K$ as well, and we use $f=22.5^{+1.3}_{-1.5}$~GeV from the UChPT results in Ref.~\cite{Fu:2021wde}.

\begin{figure}[t]
	\centering
	\includegraphics*[width=0.48\textwidth,angle=0]{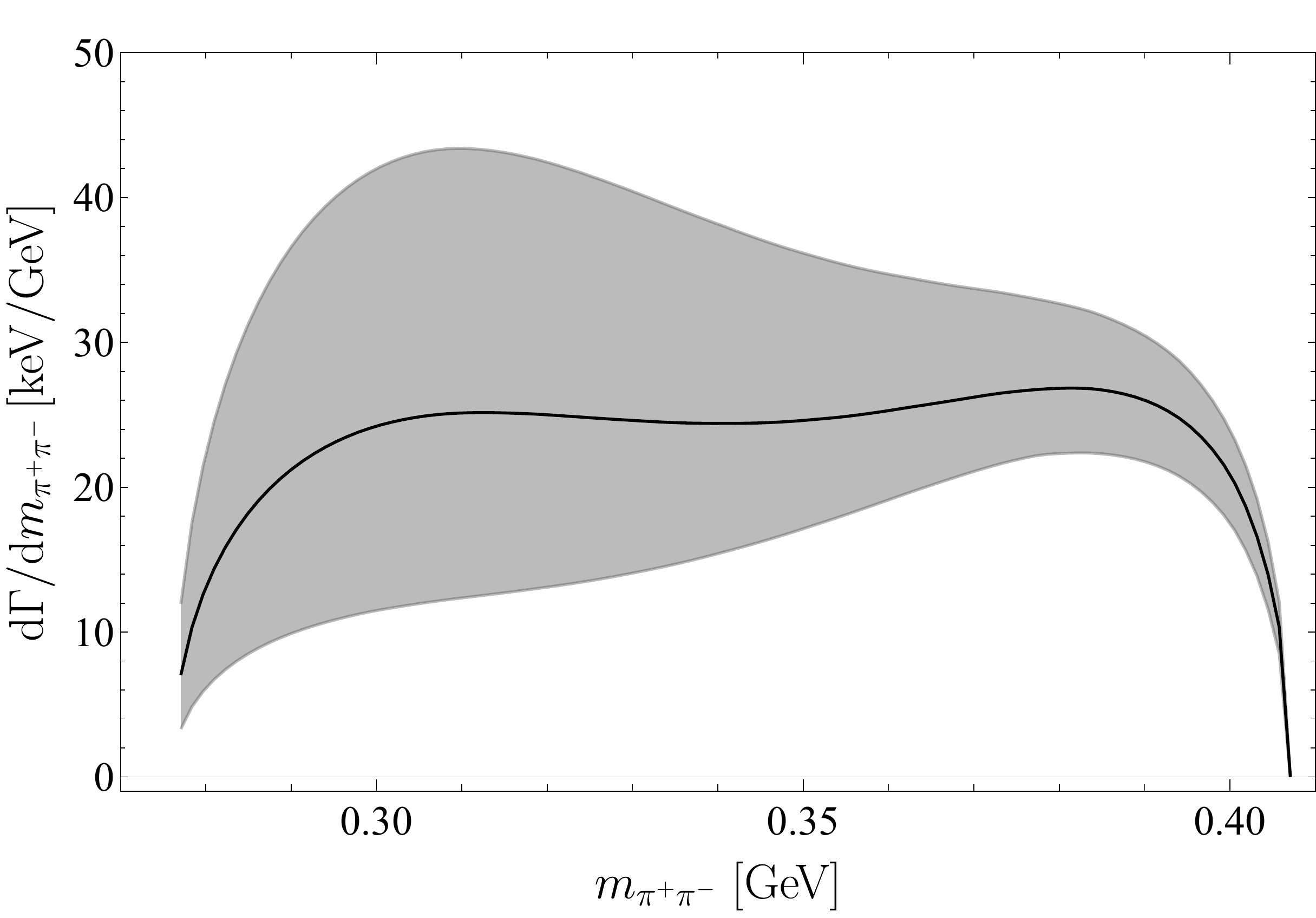}~~~~
	\includegraphics*[width=0.48\textwidth,angle=0]{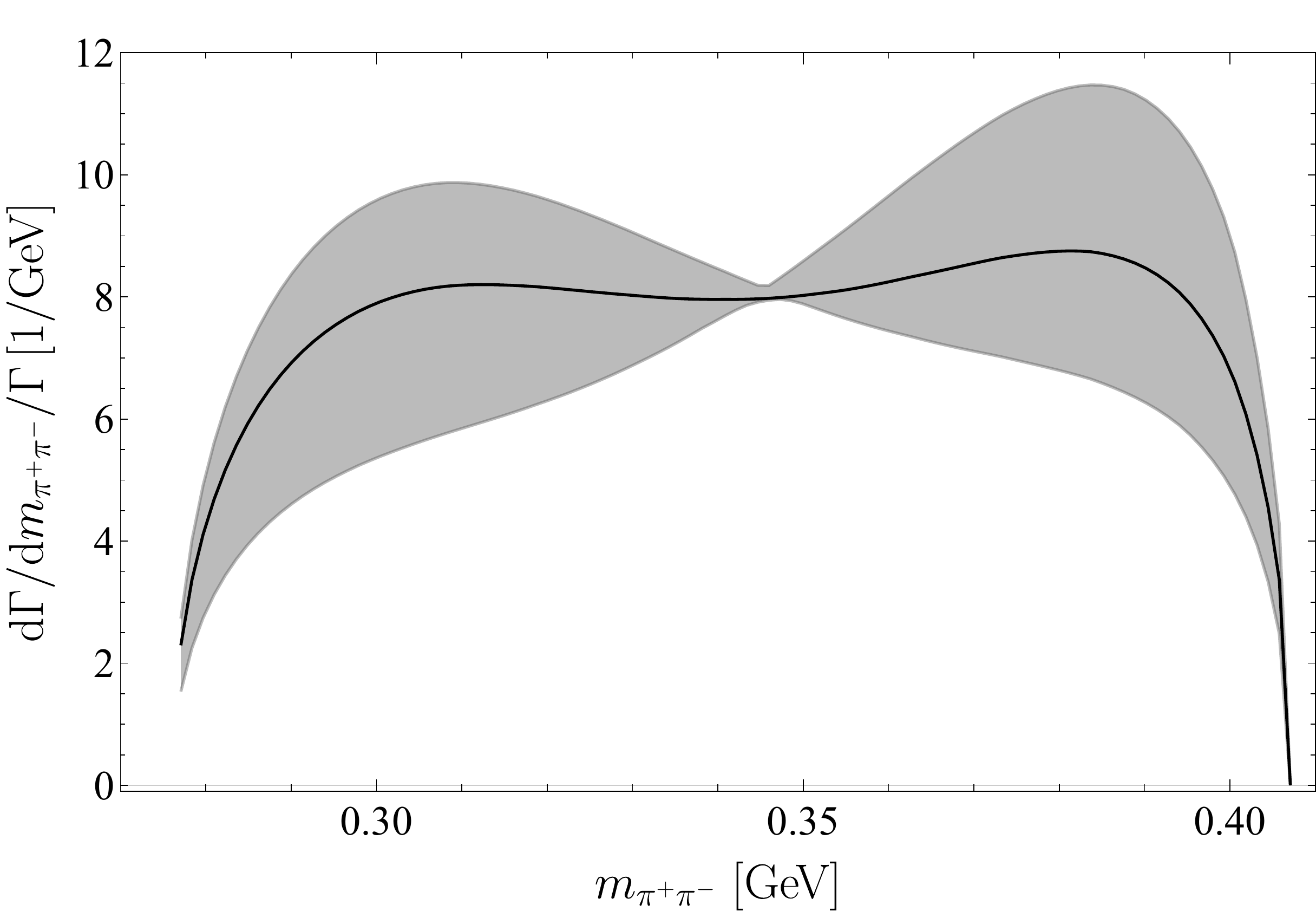}
	\caption{
		The invariant mass distributions of $\pi^+\pi^-$ for the decay of $B_{s 1}^{0} \to B_s^{0} \pi^{+} \pi^{-}$. Notations are the same as Fig.~\ref{fig:IMD-err2}.
	}
	\label{fig:IMD-err-Bs1}
\end{figure}

Taking 5774~MeV~\cite{Fu:2021wde} as the $B_{s1}$ mass, the partial width of the $B_{s 1}^{0} \to B_s^{0} \pi^{+} \pi^{-}$ is predicted to be
\begin{equation}
	\Gamma(B_{s 1}^{0} \to B_s^{0} \pi^{+} \pi^{-}) = (3\pm1)~{\rm keV}.
\end{equation}
The predicted $\pi^+\pi^-$ invariant mass distribution and the one normalized to the above partial width are shown in the left and right panels of Fig.~\ref{fig:IMD-err-Bs1}, respectively.

\section{Summary}

In this paper, we have calculated the decay width of the $\dso\to\dspipi$ under the assumption that the $\dso$ is an isoscalar $D^*K$ hadronic molecule. 
The $S$-wave $\pi\pi$ final state interaction is taken into account using a dispersive approach.
We find that the ratio of partial decays widths $\Gamma(\dso\to\dspipi)/\Gamma(\dso\to D_s^{*+}\pi^0)$ in the molecular picture agrees with the measured value, which may be regarded as a support of the $D^*K$ molecular picture for the $\dso$.
Although the decay $\dso\to\dspipi$ can proceed preserving isospin symmetry while the decay $\dso\to D_s^{*+}\pi^0$ violates isospin symmetry, the former has a smaller width due to the three-body phase space suppression.

We also find that the $\pi^+\pi^-$ invariant mass distribution of the decay $\dso\to\dspipi$  can be used to disentangle models for the $\dso$. In the $D^*K$ molecular picture, the distribution has a double bump structure, due to the $D^*K$ loop diagrams and $\pi\pi$ FSI, while in the compact state picture, in which the $D_{s1}D^*K$ coupling is negligible, the distribution has a single broad bump. 
The $\pi^+\pi^-$ invariant mass distribution can be measured at the LHCb and Belle~II experiments.

Furthermore, we also make predictions for the decay $B_{s 1}^{0} \to B_s^{0} \pi^{+} \pi^{-}$ where the $B_{s1}^0$ is the bottom partner of the $\dso$. The partial width is predicted to be $(3\pm1)$~keV. The $B_{s1}^0$ may be searched for at the LHCb experiment.

\acknowledgments 
This work is supported in part by the National Natural Science Foundation of China (NSFC) and the Deutsche Forschungsgemeinschaft (DFG) through the funds provided to the Sino-German Collaborative Research Center TRR110 “Symmetries and the Emergence of Structure in QCD” (NSFC Grant No. 12070131001, DFG Project-ID 196253076); by the Chinese Academy of Sciences (CAS) under Grant No. XDB34030000; by the NSFC under Grants No. 12125507, No. 11835015, and No. 12047503; by CAS through the President’s International
Fellowship Initiative (PIFI) (Grant No. 2018DM0034); and by the VolkswagenStiftung (Grant
No. 93562).

\bibliography{refs.bib}

\end{document}